\documentclass[aps,raggedbottom,nobalancelastpage,prb,twocolumn,10pt,amsmath]{revtex4-1}

\usepackage{graphicx}

\newcommand{\dvec}[1]{\ensuremath{\boldsymbol{#1}}}
\newcommand{\vk}{\dvec{\mathrm{k}}}

\newcommand{\erfc}{\ensuremath{\mathrm{erfc}}}

\begin{document}

\title{Compressibility of graphene}
\author{D. S. L. Abergel}
\author{E. H. Hwang}
\author{S. Das Sarma}
\affiliation{Condensed Matter Theory Center, Department of Physics,
University of Maryland, College Park, MD 20742}

\begin{abstract}
	We develop a theory for the compressibility and quantum
	capacitance of disordered monolayer and bilayer graphene including
	the full hyperbolic band structure and band gap in the latter case.
	We include the effects of disorder in our theory, which are of 
	particular importance at the carrier densities near the Dirac point.
	We account for this disorder statistically using two different
	averaging procedures: first via averaging over the density of carriers
	directly, and then via averaging in the density of states to produce
	an effective density of carriers.
	We also compare the results of these two models with experimental
	data, and to do this we introduce a model for inter-layer screening
	which predicts the size of the band gap between the low-energy
	conduction and valence bands for arbitary gate potentials applied to
	both layers of bilayer graphene.
	We find that both models for disorder give qualitatively correct
	results for gapless systems, but when there is a band gap in the
	low-energy band structure, the density of states averaging is
	incorrect and disagrees with the experimental data.
\end{abstract}

\maketitle

\section{Introduction \label{sec:intro}}

Recently, graphene has become a highly-studied electronic system with
many potential uses in electronic devices, optics, and sensing
applications.\cite{abergel-advphys} 
While significant theoretical effort has been concentrated on the transport
properties\cite{dassarma-rmp} (such as the minimal conductivity,
localization effects, and signatures of relativistic behavior of the
charge carriers), bulk thermodynamic quantities such as the
compressibility of the electron liquid in monolayer and bilayer graphene
have also received substantial attention. 
The compressibility is an important quantity because it is
possible to extract information about the electron-electron interactions
and
correlations directly from these measurements, and to gain information
about fundamental physical quantities such as the pair correlation
function. The quantum capacitance\cite{john-jap96,fang-apl91} is also an
important quantity in the context of device design and fundamental
understanding of the graphene system.
In fact, the compressibility is often inferred from the measurement of
quantum capacitance.

Experiments\cite{martin-natphys4} have shown that the
contribution to the compressibility of monolayer graphene from
electron-electron interactions is small (amounting to a $10-15\%$
renormalization of the effective Fermi velocity). This was explained
within the Hartree--Fock approximation.\cite{hwang-prl99} 
Capacitance measurements of bilayer graphene have also been
carried out recently,\cite{young-arXiv,henriksen-prb82}
and they purport to show data which indicates that the quadratic
approximation of the band structure of this material does not produce
the correct predictions for the quantum capacitance of the graphene
sheet. Other measurements which claim to access the quantum capacitance
directly have shown similar behavior.\cite{xia-natnano4}

Theoretical work has been published which considers the compressibility
of monolayer\cite{hwang-prl99} and bilayer\cite{kusminskiy-prl100}
graphene at the Hartree-Fock level and in the random phase
approxmation\cite{borghi-prb82} (RPA). 
They find that interactions are more important in the bilayer system
(due to the finite density of states at charge-neutrality), but that in
the RPA the contributions from exchange and correlation almost cancel
each other out leaving a small positive compressibility at all
densities.
A consensus has developed that many-body corrections to the
compressibility in graphene are at most of the order of 10\% at
reasonable carrier densities.\cite{sensarma-unpub}

In this article we discuss theoretically the compressibility of
monolayer and bilayer graphene, and specifically employ the
four-band model of bilayer graphene (which generates the non-quadratic
band structure). 
We will highlight the similarities and differences between this model
and the linear and quadratic cases, and take into account the
possiblility of a gap at the charge-neutrality point.
We find that the non-quadratic band structure significantly affects the
theoretical predictions for the compressibility and brings them into
qualitative and semi-quantitative agreement with experimental findings. 
We also include disorder in the form of electron--hole puddles created
by charged impurities in the vicinity of the graphene. We discuss two
different procedures for including this effect statistically and
determine the validity of each model by comparing with experimental
data. The first assumes that the disorder creates a spatial variation
in the density of carriers which can be modelled by averaging a physical
quantity such as compressiblity over a range of densities. The second
model
assumes that disorder affects the local density of states which can
be averaged to give an effective bulk density of carriers which can then
be used
to compute the physical quantities of interest. We find that for gapless
graphenes, inclusion of disorder by either method predicts qualitatively
correct results in agreement with experiments, but when a gap is opened,
the stage at which the averaging is done makes a critical difference.

Therefore, we have dual complementary theoretical goals in this work.
Our primary goal is to develop an accurate (and essentially
approximation-free)
theory for the graphene compressibility within the non-interacting
electron model which incorporates all aspects of the realistic band
structure as well as effects of the disorder within reasonable and
physically motivated approximations. Our secondary goal is to compare
our theory closely and transparently with the existing experimental data
on the compressibility of bilayer graphene so as to obtain some
well-informed conclusions about the possible role of many-body
exchange-correlation effects, which have recently been studied the the
literature\cite{kusminskiy-prl100,borghi-prb82,sensarma-unpub} but which
are beyond the scope of the current work. We believe that an accurate
calculation of graphene compressibility within the free electron theory
neglecting all interaction effects is warranted for (at least) three
reasons: (\textit{i}) without an accurate one-electron theory,
quantitative conclusions regarding many-body effects are not useful;
(\textit{ii}) many-body theory is, by definition, approximate and, as
such, subject to doubts; (\textit{iii}) many-body corrections to
graphene compressibility are claimed to be small with considerable
cancellation between exchange and correlation
effects\cite{borghi-prb82}. In comparing our theoretical results with
the experimental data, we find that the inclusion of disorder effects in
the theory is of qualitative importance, particularly at low carrier
densities near the charge-neutral Dirac point (where the many-body
effects are expected to be largest). The inclusion of disorder through
two alternative physical approximations and the eventual validation of
one of the models of disorder by comparing with the existing
experimental data is an important accomplishment of our theory.

To outline the structure of this article, we begin in the remainder of
this section by giving details of the band structures of the three types
of graphene which we consider. 
In Section \ref{sec:screening} we introduce a model for determining the
band gap in bilayer graphene with arbitrary gate potentials
applied to both layers.
Then, in Section \ref{sec:dmudn} we
describe the calculation of $K=\frac{d\mu}{dn}$ in the homogeneous
(non-disordered) situation (where $\mu$ and $n$ are the chemical
potential and carrier density respectively) for each of these three cases and discuss the
main features of this quantity in relation to the quantum capacitance
and compressibility. 
In Section \ref{sec:dismodels} we introduce the two models for the
inclusion of disorder and present the fundamental theoretical results
for each.
Then, in Section \ref{sec:expcomp} we compare the predictions of our
theory to experminental data given by capacitance measurements of
bilayer graphene which include the quantum capacitance of the 2D
electron gas.
Finally, in Section \ref{sec:conclusion} we summarize our results and
give some general discussion of our findings.

\begin{figure}[tb]
	\centering
	\includegraphics[]{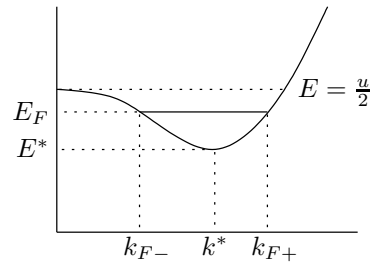}
	\caption{Sketch of the `sombrero' dispersion in hyperbolic
	graphene with $u>0$. The depth of the minimum is exagerrated for the
	purposes of illustration. The various wave vectors and energies that
	are labeled are defined in the text. \label{fig:sketch}}
\end{figure}

In the following, we discuss three different models for the
single-particle dispersion of graphene which we denote by $E_k$.
We shall refer to these three cases as linear, quadratic,
and hyperbolic graphene to distinguish them.
For the monolayer, we have $E^l_k
= \hbar v_F k$ as is well-known, and we indicate all quantities which
follow from this dispersion by attaching a superscript `$l$' as we have
here. For bilayer graphene, there are two commonly-used
approximations for the low-energy single-particle dispersion. The
simplest is the quadratic approximation $E^q_k = \hbar^2 v_F^2 k^2 /
\gamma_1 = \hbar^2 k^2 / (2m)$ where $\gamma_1$ is the inter-layer
hopping parameter from
the tight-binding theory and $m=\gamma_1/(2v_F^2)$ is the effective mass
of the electrons. This dispersion comes from a low-energy effective
theory of bilayer graphene\cite{mccann-prl96} which essentially discards
the two split bands and confines electrons to those lattice sites not
involved in the inter-layer coupling.
This quadratic bilayer dispersion is assumed to hold for low carrier
densities, \textit{i.e.} at low Fermi energy.\cite{dassarma-rmp}
Alternatively, the full tight-binding theory\cite{mccann-prb74} yields
the single-particle dispersion
\begin{equation}
	E^h_k = \pm\sqrt{ \frac{\gamma_1^2}{2} + \frac{u^2}{4} + 
	\hbar^2 v_F^2 k^2 - \sqrt{ \frac{\gamma_1^4}{4} + 
	\hbar^2 v_F^2 k^2 ( \gamma_1^2 + u^2)}} \label{eq:spdisp}
\end{equation}
for the low-energy bands, which is illustrated for the conduction band
in Fig.~\ref{fig:sketch}. 
This dispersion allows for the inclusion of a band gap parameterized by
the energy $u$ which corresponds to the potential energy difference
between the two layers. The `sombrero' shape of the dispersion
implies that the minimum gap is found at wave vector $k^\ast =
\frac{u}{2\hbar v_F} \sqrt{\frac{2 \gamma_1^2 + u^2}{\gamma_1^2 + u^2}}$
and the energy of the minimum is therefore $E^\ast=\frac{\gamma_1
u}{2\sqrt{\gamma_1^2+u^2}}$. This non-monotonic dispersion relation
leads to some important features in the Fermi surface at low density.
For medium and high density regimes (when $k_F > \frac{u}{\hbar v_F}$),
the Fermi wave vector is given by $k_F = \sqrt{\pi n}$ and the Fermi
surface is disc-shaped. However, if $\sqrt{\pi n} < \frac{u}{\hbar v_F}$
(as sketched in Fig.~\ref{fig:sketch}) then the Fermi energy is in the
`sombrero' region and the Fermi surface is
ring-shaped\cite{stauber-prb75} with $k_{F-} < k < k_{F+}$ and
\begin{equation}
	k_{F\pm} = \frac{1}{2\hbar v_F} \sqrt{ 
		\frac{\pi^2 \hbar^4 v_F^4 n^2 + \gamma_1^2 u^2}{\gamma_1^2+u^2}
		+ u^2 \pm 2\pi \hbar^2 v_F^2 n }. \label{eq:kofn}
\end{equation}
Clearly these characteristic features of the full hyperbolic bilayer
band dispersion are lost in the linear (``high-energy'') and quadratic
(``low-energy'') approximations often adopted in the literature for
simplicity.

The parameter $u$ can be treated in two ways. Initially (and in
Section \ref{sec:dismodels}), we consider it to be a phenomenological
parameter which can be chosen arbitrarily to represent a finite gap
at charge-neutrality. However, in order to accurately compare our
theory with experiment we must determine a method of relating $u$ to
the external gate potentials. To do this accurately, the screening
of the external field by the charge on the two layers of bilayer
graphene must be taken into account, and we describe this process in
Section \ref{sec:screening}. 

\section{Inter-layer screening \label{sec:screening}}

\begin{figure}[tb]
	\centering
	\includegraphics[]{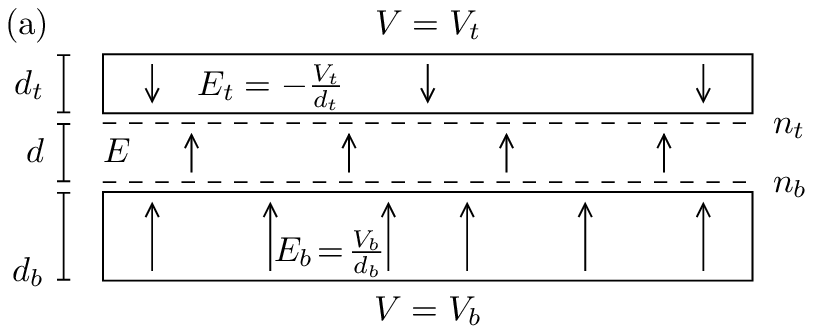}
	\includegraphics[]{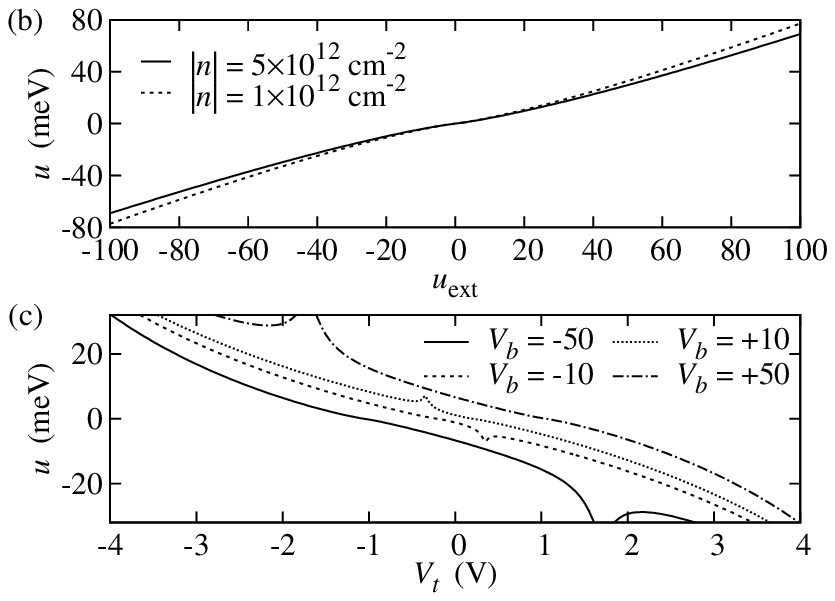}
	\caption{(a) The geometry of the screening problem. The external
	field $E_{\text{ext}}=(E_b+E_t)/2$ is screened by redistribution 
	of the charge density $n=n_b+n_t$ between the two layers of
	the graphene.
	(b) The screened gap $u$ as a function of the bare gap
	$u_\text{ext}$ for various densities.
	(c) The screened gap as a function of the top gate voltage for
	various fixed back gate voltages and the experimental parameters
	used in the experiments of Young \textit{et
	al.}\cite{young-arXiv,exppar} Back gate voltages are quoted in
	Volts.
	\label{fig:screengeom}}
\end{figure}

In order to make an accurate comparison with experimental data, we must
properly map the gate voltages applied in experiment to the
parameters $n$ and $u$ which appear in our theory. 
We assume the situation sketched in Fig.
\ref{fig:screengeom}(a).
The density is easily computed via elementary considerations to be $n=
\epsilon_0 \left( V_b \epsilon_b / d_b + V_t \epsilon_t / d_t\right) / e$
where $V$ is the voltage applied to a gate, $d$ is the width of the gate layer,
$\epsilon$ is the dielectric constant of the gate, and the subscript $b$
and $t$ respectively denote the bottom and top gates. The zeroeth-order
approximation for the on-site potential $u$ is to say that the
electric field created by the gates directly gives the electric
field $E$ in the graphene. The potential is then $u_{t,b} = \pm e E
d/2$. However, this approximation does not take into account the
screening effect of the charge on the two graphene layers.
Previously, various authors have investigated this problem and shown
that screening reduces the size of the gap
\cite{mccann-prb74,min-prb75,castro-prl99}. We extend this
work to the case where the total density is finite and gates are
present both above and below the graphene layer.

Since the screening depends only on the electric field created by
the gates and the charge densities on the two layers, it is independent
of the disorder caused by charged impurities. 
Both the screening and the disorder effects are determined by the net
carrier density, and both leave this quantity unchanged. 
Therefore, we can use the following analysis to set the size of the gap
for any given set of external conditions, and use the resulting band
structure to calculate the effect of disorder for that situation. This
is allowed because neither disorder nor the screening induces any extra
charge in the graphene. Additionally, we will assume in Section
\ref{sec:dismodels} that the disorder potential is the same in both
layers meaning that it does not induce any inter-layer scattering and
therefore leaves the net carrier density in each layer unchanged.

According to Gauss' law, the electric field directed away from an
infinite-sized, positively charged sheet is $E=qn/2\epsilon_0$, where
$q=-|e|$ is the charge of the carriers on the sheet. Hence, we can write
the internal field (i.e. the field between the graphene layers) as 
\begin{equation}
	E=E_\mathrm{ext} + \frac{e}{2\epsilon_0}(n_t-n_b).
	\label{eq:Eint}
\end{equation}
That this internal field is different from the external one causes the
free charge on the two layers to rearrange, which in turn further
modifies the internal field. A self-consistent procedure can be
constructed to find the final solution for the layer charge densities
and the internal electric field. This additional field causes an
additional Hartree energy of $u_t - u_b = edE$ to be added to the
on-site potential in the original Hamiltonian. This modified on-site
potential will be smaller than in the unscreened case, and therefore the
gap will also be smaller.

We implement this procedure for a given pair of gate voltages $V_b$ and
$V_t$ as follows. The first step is to compute the unscreened on-site 
potential in the two layers, which is given by the expression above with
$n_t = n_b$ so that $E=E_\mathrm{ext}$. This gives $u_\mathrm{ext} =
edE_\mathrm{ext} = e d (V_b/d_b - V_t/d_t)/2$. We assume that the
potential is symmetrical so that $u_t = u/2$ and $u_b = -u/2$. This
determines the single-particle Hamiltonian, from which we can compute
the wave functions. The layer density is then found by summing the wave
function weight in each layer:
\begin{equation*}
	n_t = \sum_\mathrm{states} \left( |\phi_{At}|^2 + |\phi_{Bt}|^2
	\right)
\end{equation*}
where $A$ and $B$ label the two inequivalent lattice sites in each
layer. The sum runs over all valley, spin and wave vector states.
A similar expression is given for the lower layer by substituting
$t\to b$ throughout. Computing these densities gives all the information
required to evaluate the internal electric field, from which we can find
the modified on-site potentials via Eq. \eqref{eq:Eint}. 
These potentials are then used to calculate the modified layer
densities, and this cycle is repeated until convergence is reached.

Figure \ref{fig:screengeom}(b) shows the screened gap as a function of
the external gap for various density of carriers. The density plays a
role because the more free carriers there are, the more
effectively the external field can be screened and the smaller the
screened gap is. This is shown in the figure because the magnitude of the 
gap for $|n|=5\times 10^{12}\mathrm{cm}^{-2}$ is always smaller than
that for $|n|=1\times 10^{12}\mathrm{cm}^{-2}$. Figure
\ref{fig:screengeom}(c) shows the screened gap as a function of top gate
voltage assuming that the back gate voltage is fixed. We have used the
parameters from the experiments by Young \textit{et
al.}\cite{young-arXiv,exppar}. 
The peak in each trace is caused by the approach of the system to charge
neutrality where there are no excess carriers and hence the external
electric field is unscreened. At this point, the screened gap is
equal to the bare gap. When there is a finite density of excess
carriers, the external field is screened and the gap reduced
accordingly.
As the back gate voltage is changed, the position of the charge
neutrality point shifts and the size of the gap at charge neutrality
increases. Inclusion of inter-layer screening turns out to be important
for understanding the experimental data.

\section{Calculation of $\frac{d\mu}{dn}$ \label{sec:dmudn}}

In this section, we describe the calculation of $\frac{d\mu}{dn}$
for the homogeneous (non-disordered) system. The results derived
here will be used later to compute the same quantity for a
disordered system.
The single-particle dispersions $E_k$ of monolayer and bilayer
graphene are well known (as discussed in Section \ref{sec:intro}). 
From these relations, we can compute the
total kinetic energy for excess charge carriers by summing the
energy of carriers in filled states: $\mathcal{E} =
\sum_{\text{states}} E_k n_F(k)$ where $n_F(k)$ is the Fermi
distribution function with wave vector $k=|\vk|$. Then, the chemical
potential is defined to be the change in energy with the addition of a
particle, which is expressed as $\mu = \frac{1}{\mathcal{A}}
\frac{d\mathcal{E}}{dn}$ where $n$ is the carrier density and
$\mathcal{A}$ is the area of the graphene flake. From this we
can compute $K=\frac{d\mu}{dn}$, and the quantum capacitance $C_Q$ and
compressibility $\kappa$ are then linked to this quantity via
\begin{equation}
	C_Q = \mathcal{A} e^2 \frac{dn}{d\mu}, \qquad
	\kappa^{-1} = n^2 \frac{d\mu}{dn}.
\end{equation}
The linear and quadratic band structures lead straightforwardly to the
following expressions 
\begin{equation*}
	K^l = \frac{\hbar v_F \sqrt{\pi}}{2\sqrt{n}}, \quad\text{and}\quad
	K^q = \frac{\hbar^2 v_F^2 \pi}{\gamma_1}.
\end{equation*}
The linear dispersion leads to a square-root divergence at zero
density while the quadratic version gives a constant for all densities.
It is possible to obtain an analytical expression for $K^h$ in various
limits and we present them in the Appendix.
Here, we describe a process for writing down the answer. Starting from
the expression for the total energy, the sum is evaluated by
transforming to an integral over $k$. 
The indefinite integral in question is given by
\begin{equation}
	I(k) = \frac{2 \mathcal{A}}{\pi} \int k E_{k}^h\,dk. \label{eq:Ik}
\end{equation}
To find the total energy, we write the sum over states discussed above
as an integral over the wave vector. Then, evaluation of the Fermi
function naturally leads to an expression containing $I(k)$ at the
limiting values of the wave vector for the specific value of the
density:
\begin{equation*}
	\mathcal{E} = \frac{2\mathcal{A}}{\pi}
	\int_0^\infty k E_k^h n^{}_F(k)\, dk = I(k=k_{F_+}) - I(k=k_{F_-}).
\end{equation*}
The chemical potential can be calculated by transforming the derivative
into one for the limiting $k$:
\begin{equation*}
	\mu = \frac{1}{\mathcal{A}} \frac{d\mathcal{E}}{dn}
	= \frac{1}{\mathcal{A}} \frac{dk^{}_{F_+}}{dn}
		\frac{dI(k^{}_{F_+})}{dk^{}_{F_+}}
	- \frac{1}{\mathcal{A}} \frac{dk^{}_{F_-}}{dn}
		\frac{dI(k^{}_{F_-})}{dk^{}_{F_-}}.
\end{equation*}
However, it is clear from the fundamental theorem of calculus and the 
definition of $I$ that $dI/dk = 2\mathcal{A} k E^{}_k/\pi$ so that the
second derivative is straightforward to calculate and we have
\begin{multline}
	K^h = \frac{2}{\mathcal{\pi}} \bigg\{
	\frac{d^2k^{}_{F_+}}{dn^2} k^{}_+ E^{}_{k^{}_{F_+}}
	- \frac{d^2 k^{}_{F_-}}{dn^2} k^{}_{F_-} E^{}_{k^{}_{F_-}} \\
	+ \left( \frac{dk^{}_{F_+}}{dn} \right)^2 \left( E^{}_{k^{}_{F_+}} 
	+ k^{}_{F_+} \frac{dE^{}_{k_{F_+}}}{dk^{}_{F_+}} \right) \\
	- \left(\frac{dk^{}_{F_-}}{dn}\right)^2 
		\left( E^{}_{k^{}_{F_-}} + k^{}_{F_-}
		\frac{dE_{k^{}_{F_-}}}{dk^{}_{F_-}} \right) \bigg\}. \label{eq:Kh}
\end{multline}
The remaining derivatives are simple to compute from Eqs.
\eqref{eq:spdisp} and \eqref{eq:kofn}. As described previously, when
$|E_F|<u/2$ the values of $k_\pm$ are given by Eq. \eqref{eq:kofn}.
When $|E_F|>u/2$ (which is trivially the case for $u=0$) then $k_-=0$
and $k_+=\sqrt{\pi n}$ and only the first two terms contribute to $K^h$.
This is the central result of this section, but in the Appendix we have
evaluated this expression for various limiting cases.

\begin{figure}[tb]
	\centering
	\includegraphics[]{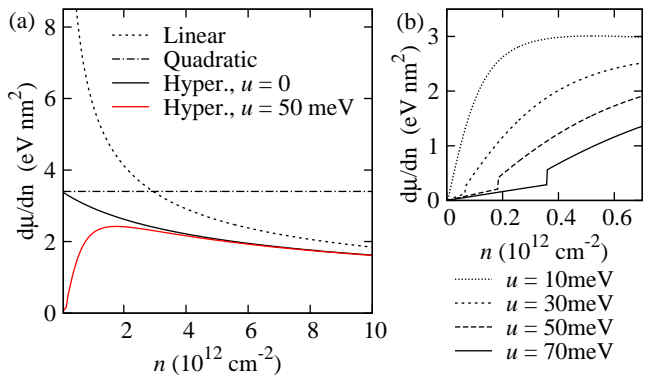}
	\caption{(a) The inverse density of states, $K=\frac{d\mu}{dn}$ for
	linear, quadratic, gapless hyperbolic and gapped hyperbolic
	($u=50\mathrm{meV}$) dispersions. 
	(b) $K$ for different gap sizes at low carrier density. 
	\label{fig:dmudn}}
\end{figure}

The three versions of $K$ are plotted in Figure \ref{fig:dmudn}(a) where
the density is assumed to be a tunable external parameter and the gap is
fixed.
In the limit of small $n$, the gapless hyperbolic band structure is
roughly quadratic and so $K^h$ is very close to $K^q$.
However, the linearity of the hyperbolic dispersion quickly asserts
itself to reduce the size of $K^h$ until, at high density, it approaches
the value of linear graphene. Thus $K^h$ in the ungapped regime
interpolates between these two limiting cases as one expects.
We also show $K^h$ for the gapped case (red or light gray line) where we
take $u=50\mathrm{meV}$. 
When the density is high and the Fermi energy is well above the gap
region, $K^h$ tends to the same functional form as for the ungapped
scenario. However, when the density
is low, $K^h$ goes to zero for any finite $u$. 
The large reduction in $K^h$ by the gap will be an important feature
for the comparison with experiment. 
In Figure \ref{fig:dmudn}(b) we show the $K^h$  at low carrier density
for several different values of the gap. One immediately notices a
discontinuity when $n=u/(\hbar v_F)$ which is caused by the change in
topology of the Fermi surface from a ring to a disc. When a gap is
present, there is also a $\delta$-function
divergence at $n=0$ (not shown in the plots) caused by the step in
the chemical potential as the Fermi energy goes from the valence
band to the conduction band.  The prefactor of this divergence is
the size of the gap so that it is not present when $u=0$.

\section{Models for disorder \label{sec:dismodels}}

We now discuss the effects of disorder caused by electron-hole puddles
which are formed by charged impurities in the
environment.\cite{martin-natphys4,rossi-prl101}
We will describe two inequivalent methods of averaging over disorder and
highlight the differences between them.
In what follows, we must be careful to distinguish global (averaged)
quantities from their local counterparts. Therefore, we use an overbar
to denote a quantity which has been averaged over disorder and which is
therefore global. We start by assuming that charged impurities (which
may be located in the substrate or near the graphene) create a local
electrostatic potential which fluctuates randomly across the surface of
the graphene sheet.  In the case of the bilayer, we assume that this
potential is felt equally by both layers. We then make the transition
from the spatial variation of the potential to an average by assuming
that the value of this potential at any given point can be described by
a statistical distribution.\cite{hwang-prb82} This distribution can then
be used to average some pertinent quantity to obtain the bulk result for
the disordered system. But it is not clear \textit{a priori} what the
optimum method of performing this average is. In order to discuss this
question, we use two different methods and (in the next section) compare
the results of each to experimental data.

The first model assumes that the disorder potential directly affects the
density of carriers in the graphene so that the density is distributed
according to a Gaussian distribution $P$ parameterized by width $\nu$
\begin{equation*}
	P(n) = \frac{1}{\nu \sqrt{2\pi}} \exp\left(-\frac{n^2}{2
	\nu^2} \right).
\end{equation*}
Then, the overall value of
$\overline{K} = \overline{\frac{d\mu}{dn}}$ can be found by evaluating
the convolution of the homogeneous (non-disordered) $K$ with this
distribution:
\begin{equation}
	\overline{\frac{d\mu}{dn}} = \int_{-\infty}^\infty
	\frac{d\mu(n-n')}{d(n-n')} P(n') dn'. \label{eq:bardmu}
\end{equation}
This has the effect of broadening any sharp features in the homogenous
$K$, such as the $\delta$-function at $n=0$ in a gapped system or the
step associated with the change in topology of the Fermi surface in the
hyperbolic band. 
Specifically, evaluating this convolution gives the following
contribution to $\overline{K}$ for the hyperbolic band from the
$\delta$-function:
\begin{equation*}
	\frac{u\gamma_1}{\sqrt{u^2+\gamma_1^2}} 
	\frac{\exp\left(-n^2/2(\delta n)^2\right)}{\delta n \sqrt{2\pi}}
\end{equation*}
so that this spike is broadened into a peak where the height is
determined by the size of the effective gap $2E^\ast$.
This model corresponds to that taken in Ref.
\onlinecite{henriksen-prb82}, except that our theory includes the
inter-layer screening which reduces the gap size.

\begin{figure}[tb]
	\centering
	\includegraphics[]{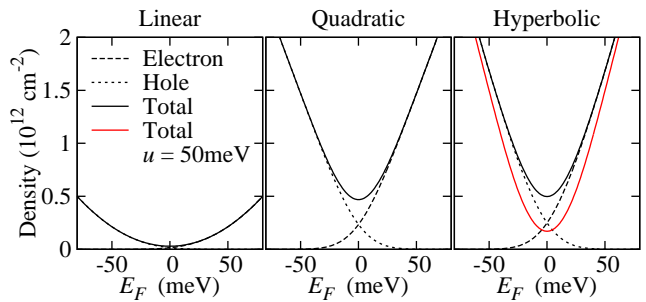}
	\caption{Electron, hole and total carrier densities for linear,
	quadratic and hyperbolic graphene for $s=20\mathrm{meV}$. 
	\label{fig:density}}
\end{figure}

The second model assumes that the disorder potential introduces
variation in the density of states (DoS). 
The potential itself is assumed to follow a Gaussian distribution $P(V)$
parameterized by width $s$, and therefore the disordered DoS is given
by\cite{hwang-prb82}
\begin{equation}
	D(E) = \int_{-\infty}^{\infty} D_0(E-V) P(V) dV \label{eq:DE}
\end{equation}
where $D_0(E)$ is the homogeneous (non-disordered) DoS.
This modified DoS then gives the zero-temperature electron density by
integrating over energy so that
\begin{equation}
	\overline{n}(E_F) = \int_{-\infty}^{\infty} D(E) \Theta(E_F-E) dE
	\label{eq:nEF}
\end{equation}
where $E_F$ is the global Fermi energy and $\Theta(x)$ is the Heaviside
step function. 
A similar expression exists for the density of holes.
This density is assumed to hold throughout the graphene so that the
averaged density can now be used to evaluate $\overline{K} 
= \frac{d\mu(\overline{n})}{dn}$. 
Since we assume that the disorder does not induce any additional charge
in the 2D system, the net carrier density $n^e-n^h$ is conserved.
Therefore, to compute the effective Fermi energy $\overline{E_F}$ in the
presence of disorder we solve the equation
$\overline{n}^e(\overline{E_F}) - \overline{n}^h(\overline{E_F}) =
n_0^e(E_F) - n_0^h(E_F)$ for $\overline{E_F}$.
However, the total carrier density $n^e + n^h$ is not conserved, and the
primary effect of disorder in this model is to introduce a finite total
density for any finite amount of charged impurities.

The integrals in Eqs. \eqref{eq:DE} and \eqref{eq:nEF} can be computed
analytically for the linear and quadratic bands if we assume that $P(V)$
is a Gaussian distribution. They are
\begin{equation*}
	\overline{n}^{l} = \frac{1}{\pi \hbar^2 v_F^2} \left[ 
		\frac{E_F^2 + s^2}{2} \erfc(\mp y) \pm \frac{E_F s}{\sqrt{2\pi}}
		e^{-y^2} \right]
\end{equation*}
and
\begin{equation*}
	\overline{n}^{q} = \frac{\gamma_1}{2\pi \hbar^2 v_F^2} \left[
	s\sqrt{\tfrac{2}{\pi}} e^{-y^2} \pm E_F \erfc(\mp y) \right].
\end{equation*}
where the upper sign corresponds to the electron density and the lower
sign to the hole density, $\erfc$ is the complementary error
function, and $y = E_F/(\sqrt{2}s)$ with $s$ the standard deviation of
the distribution of the disorder potential. 
We see that the effect of disorder is to introduce a tail in the
electron density which extends into the conduction band, the size of
which increases with increasing disorder strength. 
In the hyperbolic case, the integrals in Eqs.~\eqref{eq:DE} and
\eqref{eq:nEF} cannot be evaluated analytically, so a numerical
comparison must be done with the other two cases. 
\begin{widetext}
The expressions which must be computed for the hyperbolic band are
\begin{multline*}
	n^{he} = \frac{1}{s\sqrt{2\pi}} \frac{1}{\pi \hbar^2 v_F^2}
	\int_{-\infty}^{E_F} dE \Bigg[ \int_{-\infty}^{E-E^\ast} dV
	(E - V) \left[ 2+ \mathcal{F}(E-V) \right]
		e^{-V^2/(2s^2)} \\
	- \int_{E-\frac{u}{2}}^{E-E^\ast} dV \, (E-V) \left[ 2 -
	\mathcal{F}(E-V) \right] e^{-V^2/(2s^2)}\Bigg] 
\end{multline*}
and
\begin{multline*}
	n^{hh} = \frac{1}{s\sqrt{2\pi}} \frac{1}{\pi \hbar^2 v_F^2}
	\int_{E_F}^{\infty} dE \bigg[
	- \int_{E+E^\ast }^{\infty} dV
	(E - V) \left[ 2 + \mathcal{F}(E-V) \right] e^{-V^2/(2s^2)} \\
	+ \int_{E+E^\ast}^{E+\frac{u}{2}} (E-V) \left[ 2 - \mathcal{F}(E-V) \right]
		e^{-V^2/(2s^2)} \bigg]
\end{multline*}
\end{widetext}
where $\Delta = 1+u^2/\gamma_1^2$, 
$\mathcal{F}(x)=\Delta/\sqrt{\Delta x^2/\gamma_1^2 - \delta^2}$ and in both cases 
the second term comes from the inner Fermi surface for $E^\ast < |E| <
u/2$.

Figure \ref{fig:density} shows the density of carriers in these three
cases for a modest value of disorder ($s=20\mathrm{meV}$) and also for
$u=50\mathrm{meV}$ in hyperbolic graphene.
The small density of states in linear graphene gives a small carrier
density at low Fermi energy. Quadratic and hyperbolic graphene have
finite density of states at charge-neutrality so the density is higher
in this case.  
The density in hyperbolic graphene increases slightly quicker than
quadratic graphene because of the linear increase in the density of
states compared to the constant density of states in quadratic graphene.
Also, the finite density of states at the charge-neutrality point in
quadratic and hyperbolic graphene means that the effect of disorder is
significantly stronger in these systems than in linear graphene. In
fact, substituting $E_F=0$ into the total density gives the dependence
of the density at the charge-neutrality point as a function of the
disorder strength $s$ as
\begin{equation*}
	\overline{n}^{l}(E_F=0) = \frac{s^2}{\pi\hbar^2 v_F^2},\quad
	\overline{n}^{q}(E_F=0) = \sqrt{\frac{2}{\pi}}
		\frac{\gamma_1 s}{\pi \hbar^2 v_F^2}
\end{equation*}
so that the linear increase in $\overline{n}^q(E_F=0)$ is faster than the
quadratic one in $\bar{n}^l(E_F=0)$ for small $s$. 
The density at charge-neutrality in
hyperbolic graphene is identical to that in quadratic graphene because
the density of states is the same in both cases. We emphasize that both
models of disorder considered here are physically motivated and
\textit{a priori} there does not appear to be any practical reason to
choose one over the other. In the next section, we compare theory with
recent graphene compressibility experimental data to establish the
comparative validity of our disorder models.

\begin{figure}[tb]
	\centering
	\includegraphics[]{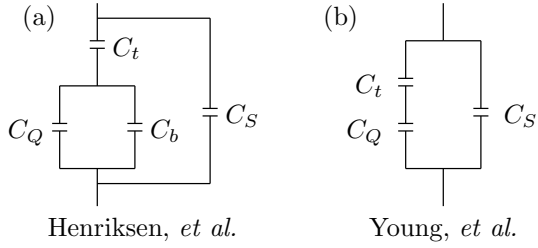}
	\caption{Classical circuit diagrams for the experiments by (a)
	Henriksen \textit{et al}.\cite{henriksen-prb82}, and (b) Young
	\textit{et al}.\cite{young-arXiv}
	\label{fig:circuits}}
\end{figure}

\section{Experimental comparison: The effect of disorder
\label{sec:expcomp}}

\begin{figure}[tb]
	\centering
	\includegraphics[]{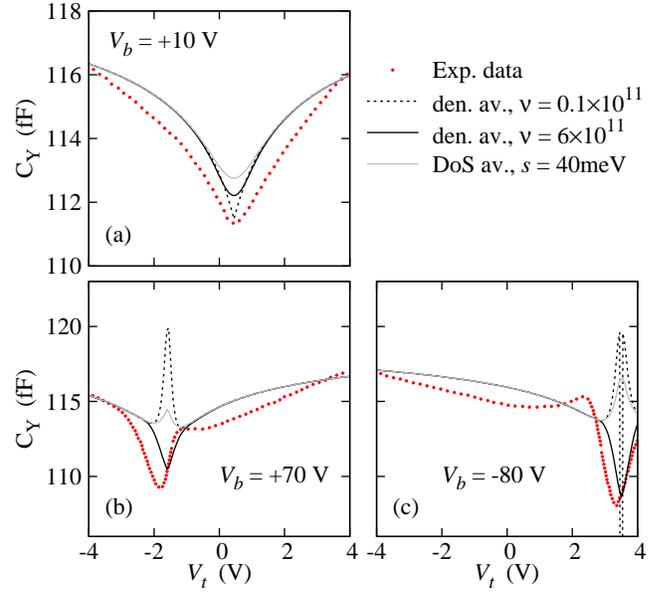}
	\caption{Comparison of theoretical results including disorder for
	the measured capacitance $C_Y$ with
	experimental data by Young \textit{et al}.\cite{young-arXiv} We have
	taken $\gamma_1 = 0.35\mathrm{eV}$, and $v_F = 1.05\times10^6
	\mathrm{ms}^{-1}$.
	\label{fig:yexpcomp}}
\end{figure}

\begin{figure}[tb]
	\centering
	\includegraphics[]{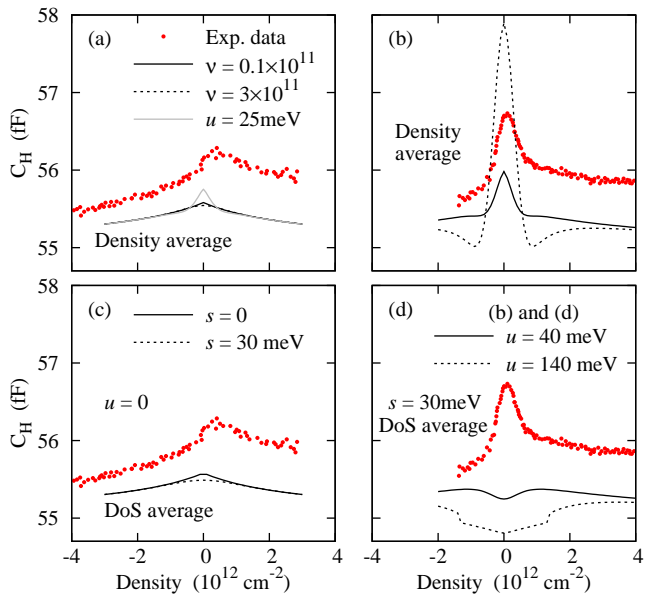}
	\caption{Comparison of theoretical results including disorder with
	experimental data by Henriksen \textit{et
	al}.\cite{henriksen-prb82} (a) $u_\mathrm{ext} = 0$ compared with
	the density averaged model for zero gap and a small gap; and
	$u=25\mathrm{meV}$ for $\nu=3\times10^{11}\mathrm{cm}^{-2}$.
	(b) $u_\mathrm{ext}$ finite, compared with the density averaged
	theory for several values of the gap. We have taken $\nu =
	3\times10^{11} \mathrm{cm}^{-2}$ and the legend is as shown in (d).
	(c) DoS averaged theory compared with $u_\mathrm{ext}=0$
	expermimental data for $s=0$ and $s=30\mathrm{meV}$.
	(d) $u_\mathrm{ext}$ finite compared with DoS averaged theory for
	two values of the gap. $s=30\mathrm{meV}$ in each case.
	We have taken $\gamma_1=0.35\mathrm{eV}$, and $v_F = 1.05\times10^6
	\mathrm{ms}^{-1}$ throughout.
	\label{fig:henexpcomp}}
\end{figure}

We now discuss our results in the context of recently published
experiments.  
We restrict ourselves to comparison with the bilayer because it is known
that the effect of disorder in monolayer graphene is small, and this has
already been discussed in the literature.\cite{hwang-prl99}
In particular, two measurements have been reported where
the capacitance of bilayer graphene has been measured and the
compressibility extracted from those measurements.
\cite{young-arXiv,henriksen-prb82} 
The quantum capacitance is linked to $\frac{d\mu}{dn}$ since $C_Q =
\mathcal{A} e^2 \frac{dn}{d\mu}$ where $\mathcal{A}$ is the surface area of the sample. The
two experiments actually measure slightly different capactitances which
are represented by the two circuit diagrams in Fig.~\ref{fig:circuits}
and correspond to the following expressions for the measured
capacitances $C_H$  and $C_Y$:
\begin{alignat*}{6}
	C_Y &= \frac{\mathcal{A}\epsilon_0 \epsilon_t}{\epsilon_t d_g + d_t} + C_S &&
	\text{Young} && \text{Ref.~\onlinecite{young-arXiv}} \\
	C_H &= \frac{\epsilon_0 \epsilon_b \epsilon_t d_g \mathcal{A}}{d_g(\epsilon_b
	d_t + \epsilon_t d_b) + d_b d_t } + C_S &\quad& \text{Henriksen} 
	& \,\,\, & \text{Ref.~\onlinecite{henriksen-prb82}}
\end{alignat*}
The notation follows that defined in Fig.~\ref{fig:screengeom}(a), 
except that $d_g$ is an effective length associated with the quantum
capacitance as
\begin{equation*}
	C_Q = \frac{\mathcal{A} \epsilon_0}{d_g} \quad\Rightarrow\quad
	d_g = \frac{\epsilon_0}{e^2} \frac{d\mu}{dn}.
\end{equation*}
The quantity $C_S$ is the background (or stray) capacitance
associated with, for example, the substrate and experimental equipment.
For relevant experimental parameters, $d_g \ll d_b, d_t$ so that $C_H
\propto d_g \propto \frac{d\mu}{dn}$. We find the relevant experimental
parameters from the papers\cite{exppar} and use them to produce the
plots shown in Figs.~\ref{fig:yexpcomp} and \ref{fig:henexpcomp}.
Note that if the quadratic band structure is used to compute $C_H$ and
$C_Y$, so that $d_g=\hbar^2 v_F^2 \epsilon_0 \pi/(e^2 \gamma_1)$ then
there is no density dependence in either $C_H$ or $C_Y$, which is
qualitatively different to the experimental data in
Figs.~\ref{fig:yexpcomp} and \ref{fig:henexpcomp}. Therefore the full
band structure is an essential part of our theory.

Figure \ref{fig:yexpcomp} shows the comparison of data from both
disorder models to the experiments\cite{young-arXiv} of Young
\textit{et al.}
In this case, the back gate is kept at a fixed potential while the top
gate is varied. This has the effect of simultaneously changing the size
of the gap and the carrier density so in this case we treat these two
quantities seperately and use the full theory of the inter-layer
screening described in Section \ref{sec:screening} to compute the size
of the effective gap.
When the gap is small at low density (\textit{i.e.} for
$V_b=10\mathrm{V}$, plot (a)) the both theories of disorder predict
qualitatively correct answers. The difference between the experimental
and theoretical curves can be
reduced by changing
the value of the stray capacitance as a fitting parameter. For very low
values of disorder, a small spike is present at zero density in the
density averaged data.
This is due to the very small band gap that is present and the
associated $\delta$-function in $\frac{d\mu}{dn}$. Larger values of
disorder smear this divergence out into a dip, the shape of which
matches the experimental data well. The DoS averaged plots do not show
this dip because carrier density is always finite in this model, and 
the effect of increasing disorder is to increase the
minimum value of the carrier density thus decreasing the depth of the
dip.
When the gap is large at zero density (\textit{i.e.} for large back 
gate voltage,
$V_b=70\mathrm{V}$, plot (b) and $V_b=-80\mathrm{V}$, plot(c)) then the
DoS averaging procedure
predicts qualitatively wrong results for low values of
disorder. There are two main features in the low density part of these
plots. First, for finite gap, Fig.~\ref{fig:dmudn} shows that
$\frac{d\mu}{dn}$ tends to zero as density decreases. This gives a peak
in the measured capacitance (because $C_Y \propto
1/(\frac{d\mu}{dn}+\frac{e^2 d_t}{\epsilon_0 \epsilon_t})$) which is blurred by finite
disorder. The second feature is the $\delta$-function at the band edge
(or, equivalently,
at zero density) which is broadened by finite disorder in the density
averaging model, but which plays no role in the DoS averaging model
because this model always gives rise to a finite density of carriers. 
This means that the DoS average cannot predict correct results in the
low-density, finite gap regime. However, the density averaging model
does include the effects of the $\delta$-function spike by sampling many
values of density, and crucially including $n=0$. This spike in
$\frac{d\mu}{dn}$ manifests as the dip in the measured capacitance and
parameters can be chosen so that the size and shape of the dip match the
experimental data quite closely. We therefore believe the
density-averaging procedure of Eq. \eqref{eq:bardmu} to be the
preferable model for including disorder over the the DoS-averaging
model.

This general picture is also seen in the comparison of the two disorder
theories with data by Henriksen \textit{et al.} as shown in
Fig.~\ref{fig:henexpcomp}. 
Panels (c) and (d) show the comparison of the experimental data to the
DoS averaged model. In this experiment, the measured capacitance is
proportional to $\frac{d\mu}{dn}$, so in the gapped case in panel (d),
the step in $\frac{d\mu}{dn}$ due to the change in topology of the Fermi
surface is clearly seen for large gap. 
The zero gap data in panel (c) shows qualitatively the correct features,
although the stray capacitance must be altered to get a better fit and
the peak is taller in the experimental data than in the theory. 
But when the gap is opened as shown in part (d), the DoS averaged 
theory predicts qualitatively wrong results because it does not account
for the broadened $\delta$-function associated with the band edges.
The experimental data also shows a distinct asymmetry between the electron and hole
density regimes. This is not explained in our model, but can be added by
allowing for an asymmetry in the underlying band structure, for example
by including specific next-nearest neighbor hops in the tight binding
model.\cite{li-prl102}
On the other hand, the density-averaging theory (shown in panels (a) and
(b)) do predict the correct behavior. For the zero-gap experiment, the
variation of the predicted $C_H$ is too small, and better agreement can
be found by allowing for a small gap to open as shown by the dashed
line. In panel (b), it is clear that a gap size of $\approx
40\mathrm{meV}$ gives a semi-quantitative fit to the experiment.

\section{Conclusion and discussion \label{sec:conclusion}}

In conclusion, we have presented a full calculation of $\frac{d\mu}{dn}$
(and hence the quantum capacitance and compressibility) for graphene and
have discussed two different methods for including disorder effects. 
These models require averaging the effect of the disorder potential in
either the density of states or directly as a variation in the local
electron density. We find that when there is no
gap in the system, these two models predict qualitatively similar
results. But when there is a gap present, the DoS averaging method fails
to account for the step in the chemical potential due to the band edges
and therefore predicts qualitatively incorrect results. 
So, we present two conclusions: First, that the single-particle theory
for the compressibility of graphenes appears to allow qualitatively
good predictions for thermodynamic quantities such as the
compressibility. Second, the inclusion of disorder via statistical
averaging must be done carefully because the choice of where the
averaging is done critically affects the outcome. We find that the
averaging should be carried out at the level of the experimental
quantity rather than the density of states.

We believe that the effects left out of the theory, particularly the
many-body exchange-correlation corrections are likely to be
qualitatively unimportant for understanding the experimental data for
currently available graphene samples which are dominated by disorder. In
particular, disorder effects are strongest at the lowest carrier
densities where exchange-correlation effects become more important
quantitatively. This makes an observation of many-body effects through
compressibility measurements quite challenging.

We thank J. P. Eisenstein, E. A. Henriksen, P. Kim, and A. Young for
discussions of their experimental results.
This work was supported by US-ONR and NRI-SWAN.

\appendix*
\section{Analytical results for $K^h$}
In this Appendix, we give the derivation of $K^h$, and write down the
full expression for it in various cases.
The full form of the indefinite integral in Eq. \eqref{eq:Ik} is
\begin{align*}
	I(k) &= \frac{2\mathcal{A}}{\pi} \int k E_{k}^h \,dk \\
	&= \frac{\mathcal{A}\gamma_1^3}{\pi\hbar^2 v_F^2} 
	\frac{1}{24\sqrt{\Delta}} 
	\bigg\{
		\sqrt{2\Delta}\left( 8x - \Phi + 1 - 4\delta^2 \right) \Xi + \\
	& \qquad\qquad + 12 \delta^2 \log\left[ 2\Delta \left(
		-\Delta+\Phi + \sqrt{2\Delta}\Xi \right)
		\right] \bigg\}
\end{align*}
where $x = (\hbar v_F k/\gamma_1)^2$, $\delta=u/(2\gamma_1)$, 
$\Phi = \sqrt{1 + 4x\Delta}$, 
$\Xi = \sqrt{ 2x - \Phi + 1 + 2\delta }$, and $\Delta = 1 + 4\delta^2$.
Explicitly differentiating with respect to $k$ verifies the Fundamental
Theorem of Calculus which we utilized in the description of Eq.
\eqref{eq:Kh}.
In the general case, the derivative of the energy with respect to the
wave vector is
\begin{equation}
	\frac{dE_{k_F}^h}{dk_F} = 
	\frac{\hbar^2 v_F^2 k_F}{E_{k_F}^h} \left( 1 - 
	\frac{\gamma_1^2 + u^2}{2 \sqrt{\frac{\gamma_1^4}{4} 
	+ \hbar^2 v_F^2 k_F^2 ( \gamma_1^2 + u^2 )}} \right). 
	\label{eq:dEdk}
\end{equation}

\subsection*{The $u=0$ limit}

In this case, $k^{}_F = \sqrt{\pi n}$ for all densities. Therefore the
derivatives of $k^{}_F$ with respect to $n$ are elementary, the
$k^{}_{F_-}$ terms in Eq.~\eqref{eq:Kh} trivially do not contribute and
we have
\begin{equation*}
	K^h_{u=0} = \frac{\gamma_1}
	{n_0 \sqrt{\frac{1}{2} + \frac{n}{n_0} - 
	\sqrt{\frac{1}{4} + \frac{n}{n_0} }} }
	\left( \frac{1}{2} -
	\frac{1}{4\sqrt{\frac{1}{4} + \frac{n}{n_0}} } \right)
\end{equation*}
where $n_0 = \gamma_1^2/(\pi \hbar^2 v_F^2)$.

\subsection*{When $u>0$, $|E^{}_F| > u/2$}

When $|E_F|>u/2$ then $k^{}_{F_+} = \sqrt{\pi n}$ and $k^{}_{F_-}=0$ 
so that the derivatives with respect to $n$ are elementary and only the
first two terms of Eq. \eqref{eq:Kh} contribute to $K^h$. Therefore
\begin{multline*}
	K^h(|E_F|>\frac{u}{2}) = \frac{\gamma_1}{2n_0\sqrt{\frac{1}{2} +
	\frac{n}{n_0} + \frac{u^2}{4\gamma_1^2} - \sqrt{\frac{1}{4} +
	\frac{n}{n_0}\frac{n_{0u}}{n_0}}}} \\
	\times \left( 1 -
	\frac{n_{0u}/{n_0}}{2\sqrt{\frac{1}{4} +
	\frac{n}{n_0}\frac{n_{0u}}{n_0}}} \right)
\end{multline*}
where $n_{0u} = (\gamma_1^2+u^2)/(\pi \hbar^2 v_F^2)$.

\subsection*{When $u>0$, $|E^{}_F| < u/2$}

However, when $|E_F|<u/2$ we need to use the expressions for the two
Fermi surfaces in Eq. \eqref{eq:kofn} so that
\begin{equation*}
	\frac{d k^{}_{F_\pm}}{dn} = \frac{\pi}{4 k^{}_{F_\pm}} \left(
	\frac{n}{n_{0u}} \pm 1 \right)
\end{equation*}
and hence
\begin{equation*}
	\frac{d^2 k^{}_{F_\pm}}{dn^2} = \frac{\pi}{4k^{}_{F_\pm}}
	\left[ \frac{1}{n_{0u}} - \frac{\pi}{4 k^2_{F_\pm}}
	\left( \frac{n}{n_{0u}} \pm 1 \right)^2 \right].
\end{equation*}
Substituting these expressions for the derivatives into Eq.
\eqref{eq:Kh} gives the simplified expression
\begin{multline*}
	K^h( |E_F|<u/2) = \frac{\pi}{8} \bigg[ \frac{1}{k^{}_{F_+}}
		\left( \frac{n}{n_{0u}}+1\right)^2
		\frac{dE^{}_{k^{}_{F_+}}}{dk^{}_{F_+}} \\
	- \frac{1}{k^{}_{F_-}} \left( \frac{n}{n_{0u}}-1\right)^2
		\frac{dE^{}_{k^{}_{F_-}}}{dk^{}_{F_-}} \bigg]
\end{multline*}
where the derivatives of the Fermi energy with respect to the Fermi wave
vector are given by making the appropriate substitutions in Eq.
\eqref{eq:dEdk}.

\end{document}